\documentclass[10pt]{article}
\newcommand{\ben}{\begin{enumerate}}
\newcommand{\een}{\end{enumerate}}
\newcommand{\be}{\begin{equation}}
\newcommand{\ee}{\end{equation}}
\newcommand{\bse}{\begin{subequation}}
\newcommand{\ese}{\end{subequation}}
\newcommand{\bea}{\begin{eqnarray}}
\newcommand{\eea}{\end{eqnarray}}
\newcommand{\bc}{\begin{center}}
\newcommand{\ec}{\end{center}}

\begin{document}
\topmargin=6em

\pagestyle{myheadings}

\title {\bf {Force in Kappa-Deformed Relativistic Dynamics}}
\author{{\bf  Piotr CZERHONIAK}\\
    {\it  Institute of Physics, Pedagogical University,}\\
    {\it  pl. S{\l}owia\'{n}ski 6, 65-069 Zielona G\'{o}ra, Poland}}
\date{}
\maketitle
\begin{abstract}
\noindent We consider the physical implications of various choices of the
three-momentum basis in the $\kappa$-deformed Poincar\'e algebra. In
particular, we find that the energy dependence of the velocity of a
$\kappa$-particle leads to unexpected features in $\kappa$-deformed
kinematics. We also discuss the notion of force in $\kappa$-deformed
dynamics, and as a tool example we investigate the motion of a
$\kappa$-deformed particle under the action of a constant force.
\end{abstract}

%-----------------------------------------------------------
%--------------- Section1 ----------------------------------
%-----------------------------------------------------------

\section{Introduction}

\noindent The possibility of description of space-time at subatomic
distances of order Planck-length ($l_p \sim 1.6\cdot 10^{-35}m$) using
the quantum deformed space-time symmetries (i.e. the Hopf algebra extension
of the Poincar\'e symmetry) is investigated since ten years
\cite{A1},\cite{LNR},\cite{MR}. This quantum deformation of the Poincar\'e
algebra and group leads us to the concept of noncommutative space-time
structure. There are some arguments that this noncommutativity of
space-time coordinates could be the effect caused by quantum gravity
\cite{A2},\cite{M}.\\
Further, we shall discuss a deformed Poincar\'e symmetry based on quantum
deformation with a dimensional (massive) parameter $\kappa$, so-called
$\kappa$-deformation (see, e.g.,\cite{LNR},\cite{MR}). This type of
deformation of the relativistic symmetry seems to be very mild because
of the following properties:

i) three dimensional nonrelativistic rotations $O(3)$ are not deformed,

ii) the energy remains the additive quantity (i.e. it has a
trivial coproduct, in the language of Hopf algebras), therefore, the
energy conservation law is still valid,

iii) the commutativity of space coordinates $x_i$ is preserved

$$
[x_i, x_j] \ = \ 0\qquad i,j=1,2,3.
\eqno(i.1)
$$

iv) the generators of the four-momentum $p_\mu$ commute

$$
[p_\mu, p_\nu] \ = \ 0\qquad \mu,\nu=0,...,3.
\eqno(i.2)
$$

v) the relativistic time coordinate $x_0$ appears to be a quantum
number and noncommuting space-time takes the form

$$
[x_i, x_0] \ = \ \frac{i\hbar}{\kappa c}x_i \ = \ il_\kappa x_i .
\eqno(i.3)
$$

\noindent where $\hbar$ - Planck constant, $c$ - velocity of light and
$l_\kappa$ describes the fundamental length (related to $\kappa$
deformation parameter) at which the time coordinate $x_0=ct$ has
to be considered as noncommutative \cite{ALN}. In ordinary quantum
mechanics (which becomes in the limit $\kappa\to\infty$) the
fundamental length $l_\kappa=0$. It is estimated that
$\kappa>10^{12} GeV$, therefore $l_\kappa <10^{-28} m$, in
practical considerations of the deformations effects one can
assume $l_\kappa\sim l_p$.\\
The commutation relation $(i.3)$ is a source of recently discussed
generalized uncertainty relations in the $\kappa$-deformed framework.
It appears that using Wigner's procedure of measurement of distances
and the relation $(i.3)$ one can conclude the existence of a "minimum
length" as a minimum uncertainty for the measurement of distances (see
also \cite{ALN},\cite{AMEL1}). This effect, in principle could be
tested experimentally in the gravity-wave interferometers (see
discussion in \cite{AMEL}).\\
The property (iv) of $\kappa$-deformed Poincar\'{e} algebra tell us that
as in nondeformed relativistic symmetry, the four-momentum generators
commute (contrary to our space-time $(i.3)$). However, taking into
account the Hopf algebra structure of three-momentum, because of
nonsymmetric coproduct, the addition law of momentum $p_i$
is more complicated (see, e.g.\cite{LNR}, [9-11]).\\
One can also consider the basis of three-momentum generators
$\tilde{p}_i$ given by a nonlinear transformation respecting $O(3)$
rotational covariance i.e.

$$
\tilde{p}_i \ = \ f\left(\frac{p_0}{\kappa c}\right) p_i
\eqno(i.4)
$$

\noindent the only physical requirement for the function $f$ is a correct
nondeformed limit i.e. $f(p_0/\kappa c)\to 1$ for $\kappa\to \infty$.
This condition follows from the Hopf algebra $\kappa$-deformation under
assumption that it is an extension of Poincar\'{e} algebra. However, the
physical consequences strongly depend on a choice of the momentum basis
i.e. the function $f$. In particular, $\kappa$-deformed particle
kinematics depends on the choice of three-momentum basis. We shall
discuss this problem in detail for the standard \cite{LNR} and
bicrossproduct basis \cite{MR}.\\
The problem how the $\kappa$-deformed particle kinematics depends on the
choice of three-momentum basis has been partially discussed in
\cite{Zakrzewski},\cite{LRZ}. In particular, the velocity of the particle
has unexpected features from the point of view of the relativistic ideas
- the velocity diminishes for large energy. The first attempt to find the
correct relativistic behaviour of $\kappa$-deformed velocity can be found
in \cite{Zakrzewski} where, by using the Poisson bracket, the monotonic
function of the velocity was found, but the generators of the
configuration space were changed. However, let us stress that within
this approach it is possible to obtain a $\kappa$-particle velocity with
physically reasonable properties. This result suggests that one can allow
to look for other momentum generators which are more acceptable from the
physical point of view.\\
In our paper we consider only the four-momentum algebra. Thus, we will not
discuss how the choice of the momentum basis changes the momentum
coalgebra, boost generators algebra or its coalgebra. However, different
momentum bases lead to different forms of Casimir i.e. the mass square
operator $M^2$ so, different dispersion relations.\\
We discuss the particle kinematics with the use the Hamilton's formalism
(see \cite{LRZ}). We start from the general $\kappa$-deformed quadratic
Casimir and use it to obtain the relations between the momentum, energy
and velocity for a particle with an arbitrary mass. We discuss it for the
two simple momentum bases (standard and bicrossproduct ones) and show
that these bases lead to unexpected properties of the velocity or
momentum. Then, using the Newtonian relation between the force
$\vec{F}$ and three-momentum, i.e. $\vec{F}=\dot{\vec{p}}$ (also
valid in the relativistic case) we derive the general formula for
the dependence of the force on the deformed velocity. This
relation shows us that the $\kappa$-deformation effects for the
moving massive particle are of order $\frac{1}{\kappa^2}$ so, it
is practically impossible to measure experimentally because of the
magnitude of $\kappa$. Our considerations suggest that the
possible $\kappa$-deformed effects can appear in very sensitive
experiments at the energies of order of $\kappa c^2$, therefore
recently discussed gamma-ray bursts observations \cite{AMEL} give
us a hope for verification or not the $\kappa$-deformed model of
space-time.

%-----------------------------------------------------------
%--------------------------- Section 2 ---------------------
%-----------------------------------------------------------

\section{Generalized $\kappa$-deformed mass condition and its
consequences for momentum and velocity}

\noindent In the $\kappa$-Poincar\'e algebra we can construct a deformed
quadratic Casimir $M^{2}$, describing the deformed mass square operator.
In this way the energy of the particle and the quadratic momentum
operator are related and this relation can be used as a starting point
for the description of the deformed particle kinematics and dynamics.
Naturally, its form depends on the particular choice of the momentum basis.
The various choices of the three-momentum basis and associated
$\kappa$-deformed Casimir can be found in \cite{LRT}, in particular the
standard basis \cite{LNR} and bicrossproduct basis \cite{MR}. In all cases
the energy generator is assumed to be additive quantity, in fact
this leads to the appearance of the term $\sinh(E/{\kappa c^2})$
in the dispersion relation. Strictly speaking, this form of the
energy dependence is related to $\kappa$-deformation.\\
If we consider the general three-momentum basis of the type $(i.4)$ then
the deformed mass condition takes the form:

\be
\left(2\kappa \sinh({E\over 2\kappa c^2})\right)^2 -{1\over c^2}f^{2}
({E\over
2\kappa c^2})\vec{p} \ ^2 = M^2.
\label{01}
\ee

\noindent where $M^2$ is the quadratic Casimir of the $\kappa$-deformed
Poincar\'{e} algebra and $f$ is an arbitrary invertible function of energy,
satisfying the boundary condition $\kappa\to\infty \Rightarrow f\to 1$ so

\be
\qquad \kappa\to\infty\qquad \Rightarrow M^2 c^4= E^2 -c^2 \vec{p} \ ^2.
\label{02}
\ee

\noindent In the rest frame ($\vec{p}=0$) we get

\be
M^2=4\kappa^2\sinh^2 \left({m_0\over 2\kappa}\right)=2\kappa^2
\left(\cosh({m_0\over \kappa})-1\right).
\label{03}
\ee

\noindent where $m_0$ is the rest mass of the particle. As was shown in
\cite{Bacry} for $\kappa \rightarrow \infty$ the relation (\ref{03}) in the
case $f=1$ is consistent with various definitions of  mass for a massive
particle which are introduced in the ordinary special relativity. Let us
notice, that the final form of (\ref{03}) is the same for the rest mass
definitions considered in \cite{Bacry}. Moreover, taking the Casimir $M^{2}$
as given in (\ref{03}) we obtain in the limit $\kappa \rightarrow \infty$
the standard mass shell condition. Therefore, for optional $f$,
eq.(\ref{01}) can be rewritten as follows
($E\geq m_0 c^2$)

\be
\cosh\left({E\over \kappa c^2}\right)- \cosh\left({m_0\over \kappa}\right)=
{1\over 2\kappa^2 c^2} f^2 \left({E\over 2\kappa c^2}\right)\vec{p} \ ^2 .
\label{04}
\ee

\noindent Assuming that the Hamilton's formalism holds for the
$\kappa$-deformed momentum
space the standard definition of the velocity

\be
v_i\equiv\dot{x}_i={\partial E\over \partial p_i} \label{05}
\ee

\noindent can be used to find the relation between the momentum and
velocity. From (\ref{01}) we obtain:

\be
p_i = {\kappa \over f^2}\left\{\sinh\left({E\over \kappa c^2}\right) -
{f'\over
f}\left[\cosh\left({E\over \kappa c^2}\right) - \cosh\left({m_0\over
\kappa}\right)\right]\right\}v_i
\label{06}
\ee

\noindent Using the quadratic momentum $\vec{p} \ ^2$ which can be easily
obtained from (\ref{04})

\be
p^2 (E) \ = \ {{2\kappa^2 c^2}\over f^2} \left(\cosh\left({E\over \kappa
c^2}\right) - \cosh\left({m_0\over \kappa}\right)\right),
\label{07}
\ee

\noindent we get the dependence of the velocity on energy

\be
v^2 \ = \ 2c^2 f^2 \left({E\over 2\kappa c^2}\right){{\cosh\left({E\over
\kappa
c^2}\right) - \cosh\left({m_0\over \kappa}\right)}\over
{\left\{\sinh\left({E\over \kappa c^2}\right) - {f'\over
f}\left[\cosh\left({E\over \kappa c^2}\right) - \cosh\left({m_0\over
\kappa}\right)\right]\right\}^2}}.
\label{08}
\ee

\noindent These energy dependencies of the momentum and velocity
will now be used for
two different functions $f$ associated with the earlier mentioned choices of
the momentum basis:\bigskip

\noindent (a) {\it the standard $\kappa$-deformed momentum
basis}  \cite{LNR}\\it corresponds to the choice $f\left(E\over 2\kappa
c^2\right) \ = \ 1$

\be
\lim_{E\to \infty} p^2 (E)\ \ = \ \lim_{E\to \infty}\left(\kappa^2 c^2
e^{E\over \kappa c^2}\right) \ = \ \infty.
\label{09}
\ee

\be
\lim_{E\to \infty}v^2 (E) \ = \  2c^2 \lim_{E\to \infty}{{\cosh\left({E\over
\kappa c^2}\right) - \cosh\left({m_0\over \kappa}\right)}\over {\sinh^2
\left({E\over \kappa c^2}\right)}} \ = \ 0.
\label{10}
\ee

\noindent The expression (\ref{10}), which is similar to the one obtained in
\cite{LNR}, means that $v(E)$ is not a monotonic energy function, but there
exists the maximum of the velocity

\be
v_{max}=c \exp(-{m_{0} \over \kappa}).
\label{11}
\ee

\noindent The region of energy for $v \geq v_{max}$ is nonrelativistic - the
velocity diminishes as energy increases (see also \cite{LRZ}). For small
energies $E<<\kappa c^2$
we can derive

\be
p^2 (E) \ \sim \ {1\over c^2}\left(E^2 - m_0^2
c^4\right)\left\{1+\frac{m_0^2}{12\kappa^2}
+\frac{1}{12}\left(\frac{E}{\kappa
c^2}\right)^2\right\} + O(\frac{1}{\kappa^4}), \label{12}
\ee

\be
v^2(E) \ \sim \ c^2\left(1 - \frac{m_0^2
c^4}{E^2}\right)\left(1+\frac{m_0^2}{12\kappa^2}\right)\left\{1-\frac{1}{4}
\left(\frac{E}{\kappa c^2}\right)^2\right\} + O(\frac{1}{\kappa^4}).
\label{13}
\ee

\noindent It is easy to estimate the value of $v_{max}$ $(11)$ for instance
for the electron $(m_e = 9.1\cdot 10^{-31} kg,
\kappa\sim 1.7\cdot 10^{-15} kg$) we get $v_{max}\sim c\cdot exp(-10^{-16})$
and it corresponds to the electron energy $E_e \sim 10^4 GeV$.\\
Because $\kappa c^2 \sim 10^{12} >> E_e\sim 1 GeV$ therefore, we can use
expansion $(13)$ to estimate an energy dependent variation in velocity
$v_0=lim_{\kappa\to\infty} v(E)$

\be
\frac{\delta v}{v} \ = \ \frac{v_0 - v}{v_0}\sim \frac{1}{8\kappa^2
c^4}\left(E^2 - m_0^2 c^4\right)\sim 10^{-26}.
\label{14}
\ee

\noindent This kind of the variation in velocity is discussed by Ellis at
al.
\cite{AMEL}.\bigskip

\noindent (b) {\it the bicrossproduct $\kappa$-deformed momentum
basis} \cite{MR}\\it corresponds to the choice $f\left({E\over 2\kappa
c^2}\right) \ = \ \exp({E\over 2\kappa c^2})$

\be
\lim_{E\to \infty} p^2 (E) \ = \ 2\kappa^2 c^2 \lim_{E\to \infty}\left[
e^{-{E\over \kappa c^2}}\left(\cosh\left({E\over \kappa c^2}\right) -
\cosh\left({m_0\over \kappa}\right)\right)\right] \ = \ \kappa^2 c^2
\label{15}
\ee

\be
\lim_{E\to \infty}v^2 (E) \ = \  2c^2 \lim_{E\to \infty}{e^{E\over \kappa
c^2}
{\left(\cosh\left({E\over \kappa c^2}\right) - \cosh\left({m_0\over
\kappa}\right)\right)}\over
{\left\{\sinh\left({E\over \kappa c^2}\right) -
\cosh\left({E\over \kappa c^2}\right) + \cosh\left({m_0\over
\kappa}\right)\right\}^2}} \ = \ \infty \label{16}
\ee

\noindent And for small energies $E<<\kappa c^2$ we get

\be
p^2 (E) \ \sim \ {1\over c^2}\left(E^2 - m_0^2 c^4\right)\left(1 +
\frac{m_0^2}{12\kappa^2}\right)e^{-\frac{E}{\kappa c^2}} +
O(\frac{1}{\kappa^4}),
\label{17}
\ee

\be
v^2(E) \ \sim \ c^2\left(1 -\frac{ m_0^2 c^4}{E^2}\right)\left(1 +
\frac{m_0^2}{12\kappa^2}\right)e^{\frac{3E}{\kappa c^2}}
+O(\frac{1}{\kappa^4}).
\label{18}
\ee

\noindent Using $(18)$ at low energies $E_e\sim 1GeV$, we find an energy
dependent variation in velocity

\be
\frac{\delta v}{v} \ \sim \ \frac{1}{2\kappa c^2}\left(3E + \frac{m_0^2
c^2}{12\kappa}\right) \sim 10^{-13}.
\label{19}
\ee

\noindent This form of the energy dependence is discussed also in
\cite{AMEL}. We see that in the case $(a)$ we get the quadratic dependence
on energy of $\delta v/v$ contrary to the linear dependence in the case
$(b)$. Therefore, the choice of the momentum basis, i.e. the function $f$,
leads to different physical properties at low energies. Also for  large
energies we observe some unconventional features in $\kappa$-deformed
kinematics. In particular, in the standard basis $(a)$ when energy grows the
velocity of the particle tends to zero $(10)$ and in the bicrossproduct
basis $(b)$ the velocity goes to $\infty$ $(16)$ and the limit of the
momentum is proportional to $\kappa c$ $(15)$.\\
It appears that the formula (\ref{01}) allows one to choose
the momentum basis which for all energies the momentum and velocity
behaviour is similar to standard relativistic one \cite{CN}. Obviously,
this choice would involve a more complicated form of the $f$ function.

%-----------------------------------------------------------
%--------------------------- Section 3 ---------------------
%-----------------------------------------------------------

\section{Force in $\kappa$-deformed dynamics}

\noindent Using the $\kappa$-deformed mass condition (\ref{01}) and
demanding that the standard relation between the force and momentum vectors
should be conserved by any deformation

\be
\vec{F} \ = \ \dot{\vec{p}},
\label{20}
\ee

\noindent we obtain

\be
\left[2\kappa^2 \sinh\left(E\over \kappa c^2\right) - {1\over
c^2}f f' {\vec{p}} \ ^2\right] \dot{E} \ = \ 2\kappa f^2
\vec{p}\vec{F}
\label{21}
\ee

\noindent or equivalently the same relation as in the case of
nondeformed relativistic dynamics

\be
\dot{E} \ = \ \vec{v}\vec{F}.
\label{22}
\ee

\noindent The force $\vec{F}$ introduced in (\ref{18}) can be
considered as a function \- $\vec{F} = \vec{F}(E,\vec{v}, \dot{\vec{v}})$
of the energy, velocity and acceleration. For simplicity, all
expressions will be derived for the standard $\kappa$-deformed
momentum basis (the ({\it a}) choice in our case).\\
From (\ref{06}) we get a simple formula

\be
p_i \ = \ \kappa \sinh\left(E\over \kappa c^2\right) v_i,
\label{23}
\ee

\noindent and therefore

\be
\vec{F} \ = \ \dot{\vec{p}} \ = \ \kappa\sinh\left({E\over \kappa
c^2}\right)\dot{\vec{v}} + {1\over c^2}\cosh\left({E\over \kappa
c^2}\right)\dot{E}\vec{v}.
\label{24}
\ee

\noindent Multiplying this expression by $\vec{v}$ and using (\ref{20})
we obtain

\be
\dot{E}\left[1 - {1\over c^2}\cosh\left(E\over \kappa
c^2\right)v^2\right] \ =
\ \kappa\sinh\left(E\over \kappa c^2\right)\dot{\vec{v}}\vec{v}.
\label{25}
\ee

\noindent Using (\ref{22}) we get the following formula

\be
\vec{F} = \kappa\sinh\left(E\over \kappa c^2\right)\left\{\dot{\vec{v}} +
{1\over c^2} \ {{\cosh\left(E\over \kappa c^2\right)(\vec{v}\dot{\vec{v}})}
\over {1 - {v^2\over c^2}\cosh\left(E\over \kappa
c^2\right)}}\vec{v}\right\}.
\label{26}
\ee

\noindent The relation (\ref{24}) in the nondeformed case
$\kappa\to\infty$ gives the standard relativistic formula

\be
\vec{F} \ = \ {E_{rel}\over c^2}\left\{\dot{\vec{v}} +
{\vec{v}\dot{\vec{v}}\over {c^2 - v^2}}\vec{v}\right\}\qquad
E_{rel}= \frac{m_{0}c^{2}}{\sqrt{1-\frac{v^2}{c^2}}}.
\label{27}
\ee

\noindent The formula (\ref{24}) can be rewritten in a more familiar
form with the use of the unit direction vector $\vec{n}$. Then
$\vec{v}=v\vec{n}$ and

\be
\vec{F} \ = \ \kappa\sinh\left(E\over \kappa
c^2\right)\left\{{\dot{v}\vec{n}\over {1 - {v^2\over c^2}\cosh\left(E\over
\kappa c^2\right)}} + v\dot{\vec{n}}\right\}.
\label{28}
\ee

\noindent This expression in the limit $\kappa\to\infty$ leads to the
well-known relativistic formula

\be
\vec{F} \ = \ {E_{rel}\over c^2}\left\{{\dot{v}\vec{n}\over {1 - {v^2\over
c^2}}} + v\dot{\vec{n}}\right\}.
\label{29}
\ee

%-----------------------------------------------------------
%--------------------------- Section 4 ---------------------
%-----------------------------------------------------------

\section{The motion of a particle under a constant force}

\noindent The simplest example of dynamics is the motion of a particle
with the rest mass $m_0$ under the action of a constant force. Fixing
the force as a constant quantity $F=m_{0}g$ (where $g$ is a constant
acceleration, for instance the gravity) with the same direction as the
initial velocity $\vec{v}$, we can discuss the consequences of the force
$\kappa$-deformation for the equations of motion.\\
It is easy to see that from (\ref{22}) a straight line motion in
the velocity direction follows. Taking this line as the $x$ coordinate, the
equation (\ref{20}) can be rewritten in a simpler form

\be
\dot{E} \ = \ vF \ = \ m_0 g v.
\label{30}
\ee

\noindent Using again the equation (\ref{10}), we get

\be
v(E) \ = \  c\frac{\sqrt{2\cosh\left(E\over
\kappa c^2\right)-2\cosh\left({m_0\over \kappa}\right)}}
{\sinh \left({E\over \kappa c^2}\right)}.
\label{31}
\ee

\noindent Therefore

\be
\dot{E} \ = \ 2m_0 gc \frac{\sqrt{\sinh^2\left(E\over
2\kappa c^2\right)-\sinh^2\left(m_0\over 2\kappa\right)}}
{\sinh\left(E\over\kappa c^2\right)}
\label{32}
\ee

\noindent and from the derivative the following relation can be obtained

\be
E(t) \ = \ 2\kappa c^2 \mbox{arcsinh}\sqrt{\left(m_0 g\over 2\kappa
c\right)^2 t^2 + \sinh^2\left(m_0\over 2\kappa\right)}.
\label{33}
\ee

\noindent Using the relation (\ref{28}) and integrating the velocity
(naturally $v=\dot{x}$) the motion of the particle is derived

\be
\cosh\left[\frac{m_0 g}{\kappa c^2}\left(x+{c^2\over g}\right)\right]
\ = \frac{1}{2}\left(\frac{m_0 g}{\kappa
c}\right)^2t^2+\cosh\left(\frac{m_0}{\kappa}\right)
\label{34}
\ee

\noindent With the assumptions that the mass of the particle is much
smaller than $\kappa$ ($m_0\leq \kappa$) and $gx\leq c^2$ we obtain the
following expansion of (\ref{32})

\be
\left(x+\frac{c^2}{g}\right)^2 \ = c^2t^2 + \frac{c^4}{g^2}\left(1 +
{1\over 12}\frac{m_0^2}{\kappa^2}\right)
\label{35}
\ee

\noindent Therefore, the last relation up to the quadratic term in
${1\over \kappa}$ has the same form as the standard relativistic
hyperbolic motion of a particle under the action of the constant
force \cite{Moller}. Because the corrections are too small, the
"deformed hyperboloid" of motion fits the standard relativistic curve
closely, with only minor departures. We see, that in order to obtain
significant differences between the deformed and standard case, the mass
of the moving particle should have the value of the order of $\kappa$.

%-----------------------------------------------------------
%---------------- Section 5---------------------------------
%-----------------------------------------------------------

\section{Closing remarks}

\noindent We showed that kinematics of the $\kappa$-deformed particle
depends on the particular choice of the three-momentum basis. The standard
and bicrossproduct momentum bases which are usually used have some
unconventional properties from the physical point of view. Therefore,
if one advocates the model of $\kappa$-deformed Poincar\'e symmetry then
the conventional relativistic notions have to be revisited or one should
find such the three-momentum basis i.e the function $f$ for which the
energy dependencies of the momentum and velocity behave similarly to the
ordinary special relativity. This problem will be considered
elsewhere (see \cite{CN}).\\
The considerations of $\kappa$-deformed motion of a particle under the
action of a constant force show that the departures of the hyperboloid
obtained in this work $(35)$ from the standard one are too small
to be observed in today's experiments. Therefore, it seems that
the dynamical behaviour of $\kappa$ - particle can not decide the
validity of $\kappa$-deformed relativistic symmetry.

\end{document}